# Essential sub-networks based on contact strength reveal folding kinetics

*Taisong Zou[1], Canan Atilgan[2], Ali Rana Atilgan[2], S. Banu Ozkan[1*]*

[1]Department of Physics and Center for Biological Physics, Arizona State University, Tempe, AZ 85287;

[2]Faculty of Engineering and Natural Sciences, Sabanci University, 34956 Istanbul, Turkey

Corresponding Author Email: Banu.Ozkan@asu.edu

ABSTRACT: It has been observed that the topology of the native state is an important determinant of protein folding kinetics and there is a significant correlation between folding rate and relative contact order (RCO) in two-state small single-domain proteins. However, as a pure topological property, RCO does not take into account residue interactions that also play an important role in folding kinetics. Using the inter-residue statistical contact potentials, we introduce weight into the residue network of contacts, and therefore define a weighted RCO. Using the weighted RCO, we can capture the folding kinetics of proteins having the same topology, but different sequence information. By constructing essential sub-networks based on the strength of the pairwise interactions, we are able to deduce the features of sequences redundant for folding events. We perform an analysis on 48 two-state proteins and the ultrafast-folding proteins, as well as mutants of the protein CI2, protein G, and protein L. Our results indicate that (i) both the weighted RCO of original residue network and that of essential sub-networks have significant correlations with the folding rate like RCO; (ii) the folding rate is critically dependent on the hydrophobic interactions for two-state folding; and (iii) the sub-networks distinguish the folding rate differences of the mutants and reveal the folding preferences of proteins.

## 1. Introduction

Recent data indicate that the topology of proteins is an important determinant of their folding mechanism.[1] Folding rate has been found to correlate with many topological properties, such as the effective chain length,[2] secondary structure length,[3] sequence-distant contacts per residue,[4] the fraction of contacts that are sequence distant[5] and the total contact distance[6] as well as effective contact order (ECO).[7,8] Additionally, ECO has been also used for exploring the folding routes and the kinetic impact of secondary structural motifs in folding.[9-11] Based on the effect of chain topology, the methods developed predict the folding rate of a protein with various degrees of success.[12,13]

In particular, Plaxco *et al*. first observed that the logarithm of in-water folding rates of two-state proteins correlate with a topological parameter named relative contact order (RCO).[14,15] RCO is the average sequence separation of a protein defined as

$$RCO = \frac{1}{L \cdot N} \sum_{}^{N} S_{ij} \qquad (1)$$

where $L$ is the length of protein, $N$ is the number of contacts in the protein, and $S_{ij}$ is the sequence separation of residues $i$ and $j$. RCO reflects the relative importance of local versus non-local contacts. The correlation between folding rates and RCO predicts that proteins with predominantly non-local interactions should fold more rapidly.

However, being a pure topological parameter, RCO does not apply to the study on mutants of the protein of interest, or different proteins with highly similar structure, but with different folding mechanisms such as protein G and protein L. Since RCO is insufficient in this respect, it is meaningful to describe a new parameter that adds the specificity of the sequences to the basic features of the RCO to study the folding mechanism. In this study, we treat the protein as a network of interacting residue pairs. We develop a revised version of RCO that incorporates specificity through local interaction strengths. We call this new parameter the weighted RCO (wRCO). We further show that it is possible to extract sub-graphs from the original network of the protein, where the wRCO still describes the folding rates. Finally, we demonstrate applications of this method to ultrafast folders and mutants, and correlate the roles of the different parts of the protein in folding mechanism to the dependence of folding rates on wRCO.

## 2. Method

**2.1. Spatial residue network:** We generate a homogeneous spatial residue network for a single protein according to its Cartesian coordinates collected in the protein data bank (PDB).[16,17] In this network, each residue is represented as a single point (node) positioned on its $C_\beta$ atom ($C_\alpha$ for Glycine residues). If any two residues are within a selected cutoff distance $R_c$, we claim there is a contact (edge) between them. Then the contact map of such a network can be represented by an adjacency matrix **A** whose elements are given by

$$A_{ij} = \begin{cases} H(R_c - R_{ij}) & i \neq j \\ 0 & i = j \end{cases} \qquad (2)$$

wherein $R_{ij}$ is the distance between residues $i$ and $j$, and $H(x)$ is the Heaviside step function whose value is 0 for $x \leq 0$ and 1 for $x > 0$. In our calculations, we locate the cutoff distance at the first coordination shell, so we take $R_c = 6.7$ Å.

**2.2. Network Parameters:** "Weighted RCO" is the modified version of RCO here. That is, we generate the weighted residue network by attributing a weight $w_{ij}$ to each contact of the homogeneous residue network. The weighted RCO is then defined as

$$wRCO = \frac{1}{L \cdot \sum w_{ij}} \sum (S_{ij} \cdot w_{ij}) \qquad (3)$$

Amongst $w_{ij}$, the energy-favored residue pairs have lower weights. Weights are attributed by using the cumulative distribution of the Thomas-Dill (TD) potential.[18] Values of the TD potential are extracted from a small data set of 37 proteins using the ENERGI method, yielding effective "potentials" of inter-residue interaction contacts. The cumulative distribution function, $f(x) = P$ [TD $\leq x$], represents the probability that the potential of a random residue pair takes on a value less than or equal to $x$ kT (Figure 1). If the TD potential value of a residue pair equals to zero, *i.e.*, neither attractive nor repulsive, we assign the weight 1 to this pair. This means that $w(0 \text{ kT}) = 1$. Then the weight of another residue pair with TD potential $x$ kT is calculated by $w(x) = f(x)/f(0)$.

It was recently shown that there is redundancy in amino acid-communication pathways, which contributes to the robustness of the protein structures.[17] In this study, our working hypothesis is that such redundancy also contributes to the protein folding kinetics. To test this idea, we utilize different screening cutoffs, $w_c$, in our network and generate two network copies. One is called "the actual sub-network," which contains all residues (nodes) and contacts (edges) weighing lower than $w_c$; the other is "the complementary sub-network", which contains all residues and contacts weighing higher than $w_c$. We term the RCO calculated from these sub-networks as aRCO and cRCO, respectively. Note that the connectivity of the chain is retained irrespective of the contact weight in all sub-networks.

## 3. Results and Discussions

**3.1. Two-state proteins.** We first study the relationship between RCO of the various protein networks and sub-networks with folding rate for 48 two-state folders (data listed in Table 1). We compute the distribution of TD contact potentials at various contacts of different sequential separation (*i.e.*, distribution of $w_{ij}$ at various $s_{ij}$) using the two-state proteins. Analysis of all contacts in these two-state folders has shown that contacts under varied sequence separations have different propensities with respect to the TD potential (Figure 2a). The region below -0.6kT belongs mostly to the hydrophobic interactions. These are accentuated in long-range contacts; i.e. $|i - j| > 4$. The skewed probability density distributions for different sequence separations give us the impetus to calculate *wRCO instead of RCO*. It has been shown that the natural logarithm of the folding rate ($\ln k_f$) is linearly dependent on RCO. The same correlation holds for wRCO (Figure 2b); thus, proteins with more local or energy-favored contacts in the residue network fold more rapidly (the correlation coefficient of $\ln k_f$ both versus wRCO and RCO is 0.75, the latter is not shown here).

As is shown above, wRCO performs as well as RCO for two-state folders. Moreover, it may provide a means to inspect which interaction types are more effective during the folding process. We use screening for this purpose. That is, given a cutoff weight $w_c$, we deduce "the actual sub-network" and "the complementary sub-network" for the original network, and compare RCO for both (called aRCO and cRCO, respectively; Figure 2c). As the cutoff is increased, aRCO correlates better with experimental folding rates, whereas the correlation of cRCO decreases. We find that there is a break-even point: The sub-network including all contacts below approximately $\varepsilon = -0.1$ kT correlates equally well with experimental rates as that constituted of all contacts above the same threshold. Including additional links above 0 kT does not improve the correlation of aRCO much, and including just these contacts deteriorates the correlations of cRCO. Interestingly, 0 kT is also the screening cut-off where the average path length of the sub-network equals to that of a protein when the average path length of various proteins are examined (*i.e.*, the average minimum number of connections that must be traversed to connect a residue pair *i* and *j* in the contact network) using the sub-networks with different cutoffs.[17] In the same manner, this might relate to the success of effective contact order (ECO) in predicting the folding rates that also takes into account the shortest path on residue network.[7,8,19] Furthermore, Figure 2c shows *significant* improvement in the -0.6 to 0 kT range for the aRCO, pointing out these as especially critical during folding events. Their exclusion both in aRCO or cRCO decreases the correlations. Along with Figure 2a, which shows that most of the interactions below -0.6 kT belong to the hydrophobic pairs, we may say that for two-state folding, the rate is critically dependent on the hydrophobic interactions. Finally, we extract all the hydrophobic contacts and form a "hydrophobic" residue interaction network (Figure 2d). Readily, wRCO of this hydrophobic network is also well correlated with folding rates. We note that the hydrophobic contacts, which include the Phe, Ile, Leu, Val, Ala, Trp, Tyr or Met pairs, take only 23 percent of all the contacts of the original network.

**3.2. Ultrafast folders.** More than a dozen proteins known so far are called ultrafast folders, due to their ability to fold on the microsecond time scale. Non-Arrhenius behavior has been observed in most ultrafast folders. That is, instead of speeding up, the folding slows down when the ultrafast folders are heated at high temperature. According to the "Thruway Search Model," the reason for non-Arrhenius kinetics is that increasing temperature expands the denatured ensemble that must be searched by the protein for it to find the downhill route to the native state.[20] At high temperature where barriers are readily overcome, this model also predicts that the ultimate speed limit to protein folding is the conformational search time spent in the denatured ensemble. In this model, the folding pathway of ultrafast folders is shown schematically as

$$T \underset{k_2'}{\overset{k_2}{\rightleftarrows}} D \underset{k_1'}{\overset{k_1}{\rightleftarrows}} N \qquad (4)$$

where *T*, *D* and *N* represent trapped states, denatured states and the native state, respectively; $k_1$, $k_1'$, $k_2$, $k_2'$ are the rate coefficients. Denatured states have direct access to the native state *N*, while trapped states do not. Thus, $k_1$ is the macroscopic transition rate from denatured states to native state and it depends on barrier crossing, whereas $k_2$ is the transition rate between trapped and denatured states. It is straightforward to show that the folding rate is given by the expression

$$k_f = k_1 \bigg/ (1 + \frac{k_2}{k_2'}) \qquad (5)$$

Usually $k_1$ dominates the folding rate at room temperature, so that the trapped states have little effect on the folding rate. Hence we see the classical Arrhenius behavior. However, at high temperature barrier crossing is no longer rate-limiting, thus $k_1$ dose not dominate the folding rate and one must take serious consideration of the effect of the trapped states as explained by in "Thruway Search Model".[20]

With the kinetic picture outlined by equations 4 and 5, we analyze the folding kinetics of ultrafast folders by taking into account the contact weights. In Figure 3, the plots of experimental folding rates versus wRCO for two situations are presented: (a) maximum folding rates extracted from thermal kinetic data (at their speed limits) and (b) folding rates at room temperature (data listed in Table 2). Here we replace the folding rates by the reciprocal of the folding times due to insufficient data on folding rates. Strikingly, near the speed limit of proteins (Figure 3a), the correlation between folding rate and wRCO is quite distinct from that at room temperature (Figure 3b). At the speed limit, the folding rate is positively correlated with wRCO, in stark contrast with the observations in two state folders and folding rates of ultrafast folders at room temperature. This positive correlation can be explained by the "Thruway Search Model." At room temperature, folding is rate-limited by traditional energy barriers, and thus the correlation between folding rate and wRCO is similar to that of general two-state proteins, leading to the negative slope in the plot of ln $k_f$ versus wRCO. However at the speed limit, the entropic search in the denatured basin governs the folding rates: The denatured basin for the conformational search is shrunk for proteins with higher wRCO due to the presence of the few favorable contacting pairs. Thus, it is harder for such proteins to go into the trapped pathway (due to the decrease in the multiplicity of favorable contacting pairs), making it easier to select the right routes to the native state, and leading to larger folding rates (positive slope in ln $k_f$ versus wRCO plots).

The current approach further enables us to study the contact types that are most important in the folding event studied, by investigating the correlation coefficients of folding rates with aRCO and cRCO at different screening cutoffs (Figures 3c, d). At the speed limit, the rates are mostly related to contacts of trapped states in the denatured basin. For a screening cutoff range of 0 kT to -0.6 kT, we find that cRCO shows as good a correlation with folding rates as wRCO (the former computed from sub-networks of chain connectivity and the retained contact pairs, the latter computed using the full residue network). This finding implicates that, since cRCO includes the complimentary contacts with energies larger than the screening cutoff, there are contacts with energy around [0, 0.1] kT that are indispensible for the routes from the trapped to the denatured states. Furthermore, aRCO up to 0.2 kT screening cutoff underperforms compared to wRCO, indicating that there are a few high energy contacts that are important for the T → D crossing. Interestingly, these are the energy ranges where the $(i, i+2)$, $(i, i+3)$ and $(i, i+4)$ contacts have enhanced probabilities of occurrence (Figure 2a).

In contrast, folding at room temperature is governed by the most cohesive interactions. We deduce this by examining aRCO as a function of screening cutoff (Figure 3d). At 0.2 kT, aRCO correlates with the experimental folding rates as well as the full wRCO. This behavior is identical to that of the two-state folders. In both cases, the rate-limiting step is due to barrier crossing related to thruway states. That the low-energy, non-local contacts are responsible for the rates of ultrafast folders at room temperature is also corroborated by cRCO (Figure 3d), where the loss of the few, lowest energy contacts already reduces the correlations.

**3.3. Mutant analysis.** The weighted residue network not only represents the topology of a native protein, but also reflects interactions between specific types of residues. Thus, wRCO can tell the differences in folding rates between mutants of the same protein, even though these mutants share highly similar topology and therefore, similar RCO value. The results obtained above on ultrafast folders also give a hint on how to perform the mutant analysis. If a secondary structural element is critical during early folding (*i.e.*, it emerges as an early event), then its behavior might be similar to the ultrafast folders near the speed limit, where the entropic search governs the formation of the secondary structure rather than barrier crossing. Therefore, we expect to see a positive correlation for cRCO versus folding rates of different mutants at that secondary structure site. On the other hand, if the secondary structure formation happens later on during folding and acts as a critical barrier crossing event, then we expect to see the behavior in two-state folders or ultrafast folders at room temperature with a negative correlation for wRCO versus folding rates. Finally, a lack of correlation between wRCO versus the folding rates of mutants at a secondary structure is expected to indicate that formation of that secondary structure is not kinetically critical for the overall folding event. We analyzed a series of two-state proteins which have significant amount of mutational data available at different secondary structural motifs. Below we explain how we compute wRCO of each mutant and analyze kinetic routes of folding events based on wRCO versus folding rates of mutants.

We determine candidate initial structures of the mutant by applying the following procedure: (i) We first change the residue type at the targeted site and save the new configuration as the raw structure of the mutant; (ii) we then utilize SANDER, a module in the Amber molecular dynamics package,[21] to relax the raw structure and obtain the corresponding minimum energy configuration; and (iii) this configuration is considered as the native structure of the mutant on which the wRCO analysis is carried out. In the current analysis, there are two sources for changes in the wRCO values upon the insertion of a mutation. Either the pair-wise interaction weights may change between the mutant residue and its neighbors, or a shift in the coordinates of the residues might occur as a result of the mutation and structure optimization leading to the addition or removal of contacts in the structure. We have performed a detailed inspection on the source of the changes in the complementary wRCO values by studying the mutants in CI2 and IgG binding domains of protein G (Table 3). The statistics show that the changes in the weights occur more often than the changes in the number of contacts, as exemplified by the mutations in Table 3. It is predominantly the change in the identity of the residue upon mutation, and not the shifts in the neighboring residues due to local conformational changes, that lead to large deviations of wRCO shown in bold in Table 3.

Figure 4a shows that the cRCO (at 0 kT cutoff) is positively correlated with the folding rate for the 20 mutants within the helix of chymotrypsin inhibitor 2 (CI2). We find that decrease in the relative importance of local contacts compared to non-local ones increases the folding rate. This unusual behavior, *i.e.*, the positive slope of the ln $k_f$ versus cRCO plot, indicates that the early formation of the helix is critical in the CI2 folding reaction (Figure 4b), a finding that has also been corroborated by experiments.[22] In our opinion, higher RCO in the complementary network (cRCO) implies that the alternative contacts that compete with the formation of the helix will be less likely since they involve long-range interactions (*i.e.*, contacts that are further separated in the sequence). Thus, as the wRCO gets lower for the set of mutants involving the residues of helix, the alternative routes competing with the formation of helix increases in number, and the folding rate decreases.

The same analysis applies to the IgG binding domains of protein G (Figure 5). In Figure 5a and 5b, the cRCO is positively correlated with the folding rate for mutants within the C-terminal hairpin and the helix of protein G, which suggests that both parts are important for folding and they emerge early in the folding. In Figure 5a, we notice that the DA47 mutation removes a buried salt bridge with Lys50 across the turn. Hence, we ignore it when calculating the correlation coefficient since our initial structure calculation methodology is not valid for the occurrence of such changes that might lead to large structural differences. Our finding of the early formation of the helix and C-terminal hairpin is consistent with results of simulations[23,24] and experiments[25,26]. Φ-value analysis of protein G indicates that the C-terminal hairpin in transition state resemble the native state much more than the denatured state, while the N-terminal hairpin and the helix are relatively disordered.[27] However, the kinetic consequences of mutation in the helix are complicated and suggest that the helix's C-terminus is better formed than the rest of the structure.[27] We also observe a negative correlation in the plot of ln $k_f$ vs wRCO for the mutants of the N-terminal hairpin which also shows that the folding of the N-terminal is still kinetically important for Protein G (Figure 5c).

When we apply our analysis to the B1 domain of protein L (Figure 6), which shares the same topology with Protein G but has a different sequence, we find a dissimilar behavior. We observe negative correlations between folding rate and wRCO for mutants within the helix, C-terminal and N-terminal hairpins with correlation coefficients 0.62, 0.61 and 0.47, respectively. There are some studies, which show that the N-terminal hairpin is largely formed in transition state, whereas the C-terminal hairpin and helix are disrupted.[28-30] However, we do not observe any positive correlations for the mutants of N-terminal hairpin, which indicates that early emergence of any secondary structure is not as pronounced as in the case of CI2 (helix) or Protein G (C-terminal hairpin). It has been shown that the C-terminal hairpin of Protein G can be stable even by itself.[31,32] Likewise, the helix of CI2 is almost fully structured and even stabilized with some tertiary interactions.[22] Yet, there are no data supporting the fully structured stable from of N-terminal hairpin of Protein L by itself in solution. Therefore, due to the sensitivity of our weighted contact analysis, it might be the case that we can only observe positive correlation (*i.e.*, signature of early emergence of secondary structure) between the folding rates and cRCO when that secondary structure is almost fully folded and stabilized in the early stages of folding.

## 4. Conclusions

We have developed the weighted relative contact order (wRCO) that combines the specificity of residues and the topology of the protein to study the folding mechanism. This new parameter fits well with the experimental data of 48 two-state folders, and shows that the folding rate is critically dependent on the hydrophobic interactions. The definition of wRCO enables one to study the folding kinetics of proteins having the same topology, but different sequence information. We further find that it is possible to deduce sub-networks of proteins created by screening a set of pairwise interactions that lead to the same relationship with the folding rates as the overall proteins.

Using this new parameter, we are able to study the folding kinetics of mutants and determine whether a secondary structural element is critical in early folding or not. Furthermore, when we apply these ideas to ultrafast folders, it is possible to distinguish the different folding mechanisms observed at the speed limit or room temperature. For ultrafast folders at the speed limit, we find that interactions with weights greater than 0 kT, and therefore statistically shorter range, play an important role in the early folding events that lead from the trapped to the denatured state ensemble.

Our findings are also relevant to the success in the foldability of the library of artificial WW domain sequences that are generated with coupled conservation, using statistical coupling analysis (SCA).[33] SCA suggests that all the information required for specifying the fold and characteristic function of a protein family may be sufficiently encoded in a small set of amino acid interactions revealed by the co-evolution analysis.[34] As turns out to be the case, sequences that have only the site-independent conservation information, are not foldable. Our results indicate that there are sub-networks that characterize the folding equally well as the whole structure points to the presence of pairs of interacting residues, which are critical for forming the correct overall topology. This further indicates the possibility of using the newly defined parameter, wRCO, in designing new sequences of a given fold.

**ACKNOWLEDGMENT.** We thank Kingshuk Ghosh for his helpful comments. Computer time was provided by the Fulton High Performance Computing Initiative at Arizona State University.

**Table 1.** Experimental folding rates[12], RCO and wRCO of two state folders.

| PDB code | ln kf | RCO(%) | wRCO(%) |
|---|---|---|---|
| 2ABD | 6.55 | 14.60 | 11.21 |
| 1LMB | 8.5 | 9.96 | 8.27 |
| 1IMQ | 7.31 | 13.80 | 12.40 |
| 2PDD | 9.8 | 12.86 | 10.03 |
| 1HRC | 8.76 | 12.80 | 11.11 |
| 1YCC | 9.62 | 13.29 | 11.73 |
| 256B | 12.2 | 8.51 | 6.32 |
| 1VII | 11.52 | 11.38 | 10.59 |
| 1BDD | 11.75 | 10.81 | 9.23 |
| 1ENH | 10.53 | 9.83 | 9.73 |
| 1EBD | 9.68 | 14.95 | 12.24 |
| 1NYF | 4.54 | 19.69 | 15.09 |
| 1PKS | -1.05 | 20.26 | 19.13 |
| 1SHG | 1.41 | 21.25 | 17.27 |
| 1SRL | 4.04 | 20.03 | 16.06 |
| 1TEN | 1.06 | 20.76 | 19.11 |
| 1WIT | 0.41 | 20.87 | 19.84 |
| 1CSP | 6.98 | 17.57 | 15.80 |
| 1MJC | 5.24 | 17.00 | 15.05 |
| 2AIT | 4.2 | 20.83 | 18.43 |
| 1PNJ | -1.1 | 16.11 | 15.39 |
| 1SHF | 4.5 | 20.91 | 16.54 |
| 1C9O | 7.2 | 16.89 | 16.10 |
| 1G6P | 6.3 | 18.91 | 16.41 |
| 1LOP | 6.6 | 16.77 | 14.13 |
| 1PIN | 9.5 | 17.61 | 17.14 |
| 1C8C | 6.91 | 13.89 | 11.22 |
| 1APS | -1.48 | 19.94 | 18.21 |
| 1HDN | 2.7 | 18.87 | 17.32 |
| 1URN | 5.73 | 17.08 | 15.65 |
| 2HQI | 0.18 | 20.12 | 18.34 |
| 1PBA | 6.8 | 16.71 | 14.48 |
| 1UBQ | 7.33 | 14.23 | 13.78 |
| 2PTL | 4.1 | 15.59 | 12.74 |
| 1FKB | 1.46 | 17.97 | 15.46 |
| 1COA | 3.87 | 17.73 | 16.31 |
| 1DIV | 6.58 | 13.40 | 11.62 |
| 2VIK | 6.8 | 12.38 | 11.73 |
| 1CIS | 3.87 | 19.06 | 17.24 |
| 1PCA | 6.8 | 13.44 | 10.93 |
| 1HZ6 | 4.1 | 14.29 | 10.98 |
| 1PGB | 6 | 15.50 | 13.66 |
| 2CI2 | 3.9 | 17.55 | 15.91 |
| 1AYE | 6.8 | 13.44 | 12.33 |
| 1RIS | 5.9 | 16.98 | 15.64 |
| 1POH | 2.7 | 18.38 | 16.77 |
| 1BRS | 3.4 | 11.47 | 10.24 |
| 2ACY | 0.92 | 19.71 | 18.10 |

**Table 2.** Folding times at speed limit[20] and at room temperature[35], wRCO and RCO of ultrafast folding proteins

| PDB code | τ at speed limit (µs) | τ at room temperature limit (µs) | wRCO(%) | RCO(%) |
|---|---|---|---|---|
| 1VII | 5.37 | 8 | 10.59 | 11.38 |
| 1PIN | 3.88 | 85 | 17.14 | 17.61 |
| 1E0L | 4.62 | 30 | 11.91 | 14.94 |
| 2PDD | 4.22 | 62 | 10.03 | 12.86 |
| 1PRB | 6 | N/A | 9.24 | 11.75 |
| 1ENH | 4.66 | 27 | 9.73 | 9.83 |
| 2A3D | 5.36 | 3 | 7.38 | 9.00 |
| 1BDC | 5.47 | 8 | 9.23 | 10.57 |

**Table 3.** The contacts that lead to a change of wRCO for some mutants of protein G and CI2, compared with the wild-type. Mutants with large deviation of wRCO are shown in bold.

| Mutants | ΔwRCO (%) | Contacts changing weight | Removed contacts | New contacts |
|---|---|---|---|---|
| *Protein G* | | | | |
| TA11 | 0.050 | (9,11), (10,11), (11,12) | | |
| **EA15** | **-0.754** | **(5,15), (6,15), (14,15), (15,16)** | **(4,50)** | **(4,15), (7,15)** |
| TA16 | -0.024 | (5,16), (15,16), (16,17), (16,33) | (3,17) | |
| TA18 | -0.062 | (3,18), (17,18), (18,19), (18,20), (18,29), (18,30) | (3,17), (18,26) | |
| AG20 | -0.053 | (1,20), (3,20), (18,20), (19,20), (20,21), (20,25), (20,26) | (20,22), (20,29) | |
| **TA25** | **0.557** | **(20,25), (21,25), (22,25), (24,25), (25,26), (25,28), (25,29)** | **(1,21), (23,45)** | **(9,55)** |
| **AG26** | **-0.480** | **(3,26), (18,26), (20,26), (23,26), (25,26), (26,27), (26,29), (26,30)** | **(8,53)** | **(26,28)** |
| **KG28** | **0.513** | **(25,28), (27,28), (28,29), (28,31), (28,32)** | | **(9,55), (28,30)** |
| AG34 | -0.093 | (31,34), (33,34), (34,35), (34,37), (34,39), (34,40), (34,54) | (7,34), (23,45), (30,34), (40,56) | (8,56), (34,36), (34,38), (41,43), (41,54), (42,44) |
| NG35 | -0.131 | (31,35), (32,35), (34,35), (35,36), (35,38), (35,40) | | (35,37), (35,39) |
| YL45 | -0.074 | (23,45), (44,45), (45,46), (45,47), (45,52) | | |
| DA47 | 0.108 | (45,47), (46,47), (47,48) | | |
| **TA49** | **0.655** | **(46,49), (48,49), (49,50), (49,51)** | | **(9,55)** |
| **TA53** | **0.682** | **(6,53), (8,53), (44,53), (51,53), (52,53), (53,54), (53,55)** | | **(9,55)** |
| **VA54** | **1.024** | **(7,54), (34,54), (39,54), (43,54), (53,54), (54,55)** | | **(5,54), (7,15), (9,55), (52,54)** |
| *CI2* | | | | |
| SA12 | -0.051 | (11,12), (12,13), (12,14), (12,15), (12,55), (12,56) | | |
| **AG16** | **-0.648** | **(8,16), (11,16), (13,16), (15,16), (16,17), (16,19), (16,20), (16,49), (16,57)** | **(16,61), (31,55)** | **(8,15), (14,16), (15,19), (16,18), (32,48), (51,58), (56,58)** |
| KG17 | 0.017 | (14,17), (16,17), (17,18), (17,20), (17,21), (17,29) | (13,17) | (17,19), (18,22), (30,49), (32,48) |
| **LG21** | **-0.453** | **(18,21), (20,21), (21,22), (21,25), (21,27)** | **(17,21), (43,64)** | **(15,19), (21,23), (21,24), (32,38)** |
| DA23 | 0.094 | (2,23), (19,23), (20,23), (22,23), (23,24) | (2,4) | (5,7), (15,19) |
| ED14 | -0.017 | (12,14), (13,14), (14,15), (14,17), (14,18) | | |
| EN14 | -0.056 | (12,14), (13,14), (14,15), (14,17), (14,18) | (43,64) | (15,19), (32,38), (32,48) |
| EQ14 | -0.045 | (12,14), (13,14), (14,15), (14,17), (14,18) | | |
| **EQ15** | **-0.436** | **(11,15), (12,15), (14,15), (15,16), (15,18)** | **(31,55), (43,64), (49,51)** | **(8,15), (15,19), (18,22), (32,38), (32,48)** |
| IV20 | -0.034 | (5,20), (16,20), (17,20), (19,20), (20,21), (20,23), (20,24), (20,27), (20,29), (20,47) | | (15,19), (32,38), (32,48) |

**REFERENCES**

(1) Baker, D. *Nature* **2000**, *405*, 39-42.

(2) Ivankov, D. N.; Finkelstein, A. V. *Proceedings of the National Academy of Sciences of the United States of America* **2004**, *101*, 8942-8944.

(3) Huang, J. T.; Cheng, J. P.; Chen, H. *Proteins-Structure Function and Bioinformatics* **2007**, *67*, 12-17.

(4) Gromiha, M. M.; Selvaraj, S. *Journal of Molecular Biology* **2001**, *310*, 27-32.

(5) Mirny, L.; Shakhnovich, E. *Annual Review of Biophysics and Biomolecular Structure* **2001**, *30*, 361-396.

(6) Zhou, H. Y.; Zhou, Y. Q. *Biophysical Journal* **2002**, *82*, 458-463.

(7) Dill, K. A.; Fiebig, K. M.; Chan, H. S. *Proceedings of the National Academy of Sciences of the United States of America* **1993**, *90*, 1942-1946.

(8) Fiebig, K. M.; Dill, K. A. *Journal of Chemical Physics* **1993**, *98*, 3475-3487.

(9) Weikl, T. R. *Proteins-Structure Function and Bioinformatics* **2005**, *60*, 701-711.

(10) Weikl, T. R.; Palassini, M.; Dill, K. A. *Protein Science* **2004**, *13*, 822-829.

(11) Merlo, C.; Dill, K. A.; Weikl, T. R. *Proceedings of the National Academy of Sciences of the United States of America* **2005**, *102*, 10171-10175.

(12) Gromiha, M. M.; Thangakani, A. M.; Selvaraj, S. *Nucleic Acids Research* **2006**, *34*, W70-W74.

(13) Zheng, O. Y.; Jie, L. *Protein Science* **2008**, *17*, 1256-1263.

(14) Plaxco, K. W.; Simons, K. T.; Baker, D. *Journal of Molecular Biology* **1998**, *277*, 985-994.

(15) Plaxco, K. W.; Simons, K. T.; Ruczinski, I.; David, B. *Biochemistry* **2000**, *39*, 11177-11183.

(16) Atilgan, A. R.; Akan, P.; Baysal, C. *Biophysical Journal* **2004**, *86*, 85-91.

(17) Atilgan, A. R.; Turgut, D.; Atilgan, C. *Biophysical Journal* **2007**, *92*, 3052-3062.

(18) Thomas, P. D.; Dill, K. A. *Proceedings of the National Academy of Sciences of the United States of America* **1996**, *93*, 11628-11633.

(19) Weikl, T. R.; Dill, K. A. *Journal of Molecular Biology* **2003**, *329*, 585-598.

(20) Ghosh, K.; Ozkan, S. B.; Dill, K. A. *Journal of the American Chemical Society* **2007**, *129*, 11920-11927.

(21) D.A. Case, *et al*. *Amber 9* **2006**.

(22) Itzhaki, L. S.; Otzen, D. E.; Fersht, A. R. *Journal of Molecular Biology* **1995**, *254*, 260-288.

(23) Kmiecik, S.; Kolinski, A. *Biophysical Journal* **2008**, *94*, 726-736.

(24) Sheinerman, F. B.; Brooks, C. L. *Proteins-Structure Function and Genetics* **1997**, *29*, 193-202.

(25) Kuszewski, J.; Clore, G. M.; Gronenborn, A. M. *Protein Science* **1994**, *3*, 1945-1952.

(26) Frank, M. K.; Clore, G. M.; Gronenborn, A. M. *Protein Science* **1995**, *4*, 2605-2615.

(27) McCallister, E. L.; Alm, E.; Baker, D. *Nature Structural Biology* **2000**, *7*, 669-673.

(28) Gu, H. D.; Kim, D.; Baker, D. *Journal of Molecular Biology* **1997**, *274*, 588-596.

(29) Kim, D. E.; Fisher, C.; Baker, D. *Journal of Molecular Biology* **2000**, *298*, 971-984.

(30) Karanicolas, J.; Brooks, C. L. *Protein Science* **2002**, *11*, 2351-2361.

(31) Blanco, F. J.; Rivas, G.; Serrano, L. *Nature Structural Biology* **1994**, *1*, 584-590.

(32) Munoz, V.; Thompson, P. A.; Hofrichter, J.; Eaton, W. A. *Nature* **1997**, *390*, 196-199.

(33) Socolich, M.; Lockless, S. W.; Russ, W. P.; Lee, H.; Gardner, K. H.; Ranganathan, R. *Nature* **2005**, *437*, 512-518.

(34) Lockless, S. W.; Ranganathan, R. *Science* **1999**, *286*, 295-299.

(35) Kubelka, J.; Hofrichter, J.; Eaton, W. A. *Current Opinion in Structural Biology* **2004**, *14*, 76-88.

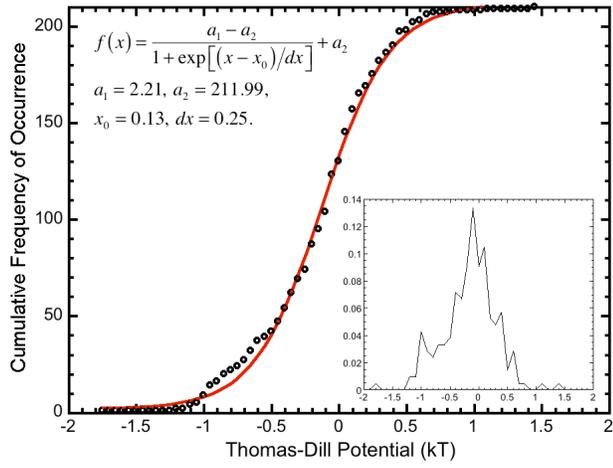

**Figure 1.** Cumulative distribution of TD potentials. The red line is the best fitted curve. The inset shows probability density distribution of TD potentials.

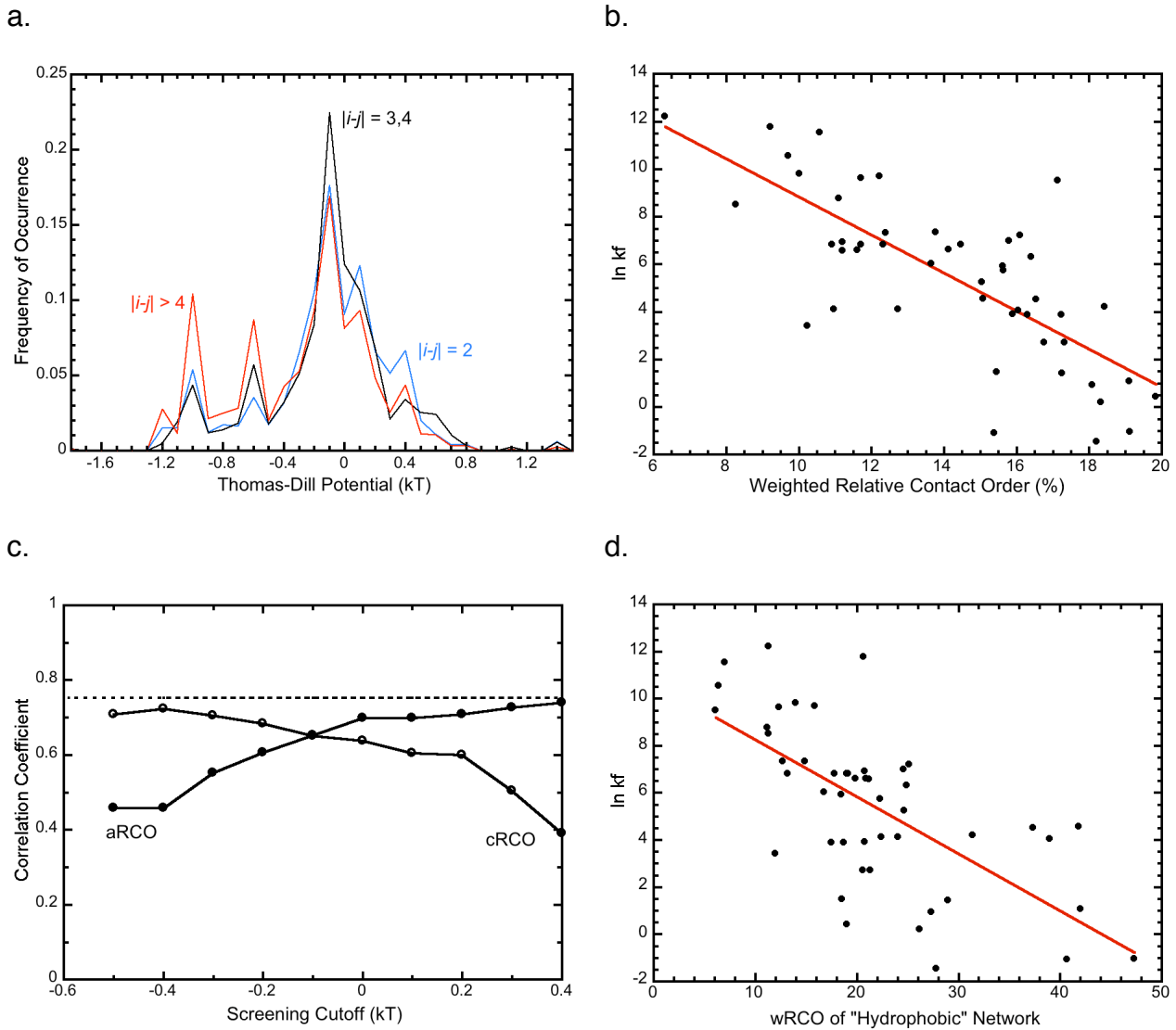

**Figure 2.** (a) Probability density distribution of TD potentials for varied Sij's. (b) ln kf vs. wRCO for two-state folders (R=0.75. Data listed in Table 1). (c) The variation in correlation coefficient with screening cutoffs. Dashed line is the wRCO value without cutoffs. (d) ln kf vs. wRCO of hydrophobic network for two-state folders (R=0.67)

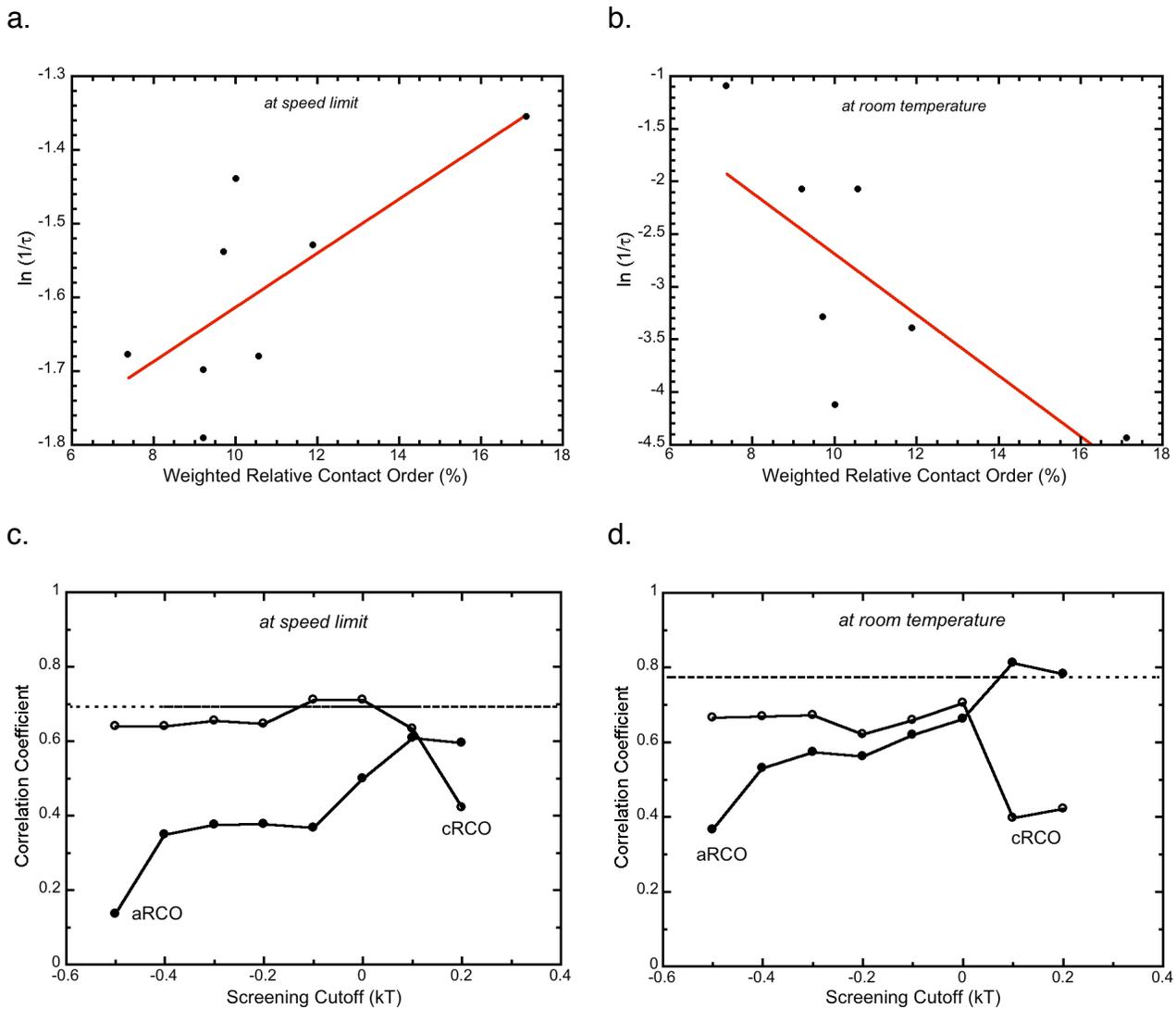

**Figure 3**. (a) ln(1/τ) vs. wRCO for ultrafast folders at speed limit (R=0.73. Data listed in Table 2). (b) ln(1/τ) vs. wRCO for ultrafast folders at room temperature (R=0.74. Data listed in Table 2). (c) The variation in correlation coefficient with screening cutoffs for ultrafast folders at speed limit. Dashed line is that from the RCO. (d) The variation in correlation coefficient with screening cutoffs for ultrafast folders at speed limit. Dashed line is that from the RCO.

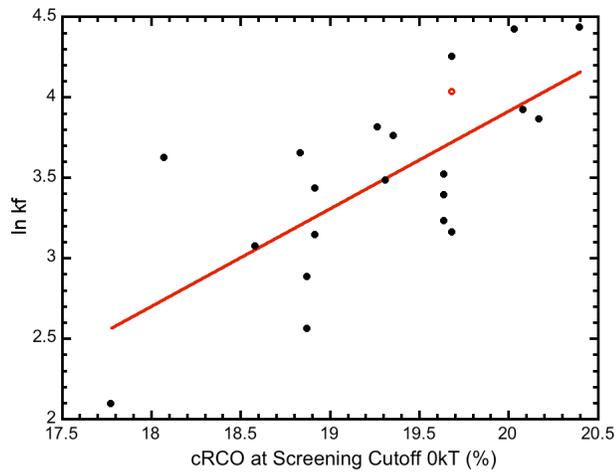 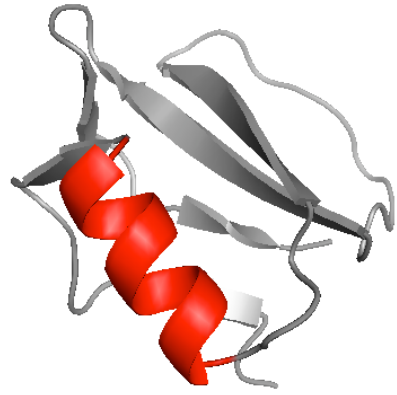

**Figure 4**. (a) cRCO at 0 kT is significantly correlated with ln(k) for the mutants within the helix of chymotrypsin inhibitor 2 (R=0.71). The red open circle marks the position of the wildtype. (b) The structure of CI2. The part which is critical for folding marked red (early formation; positive slope in ln(kf) vs cRCO plot) or blue (negative positive slope in ln(kf) vs wRCO plot). Parts that lack correlation are colored grey. This color formalism applies to figures 4-6.

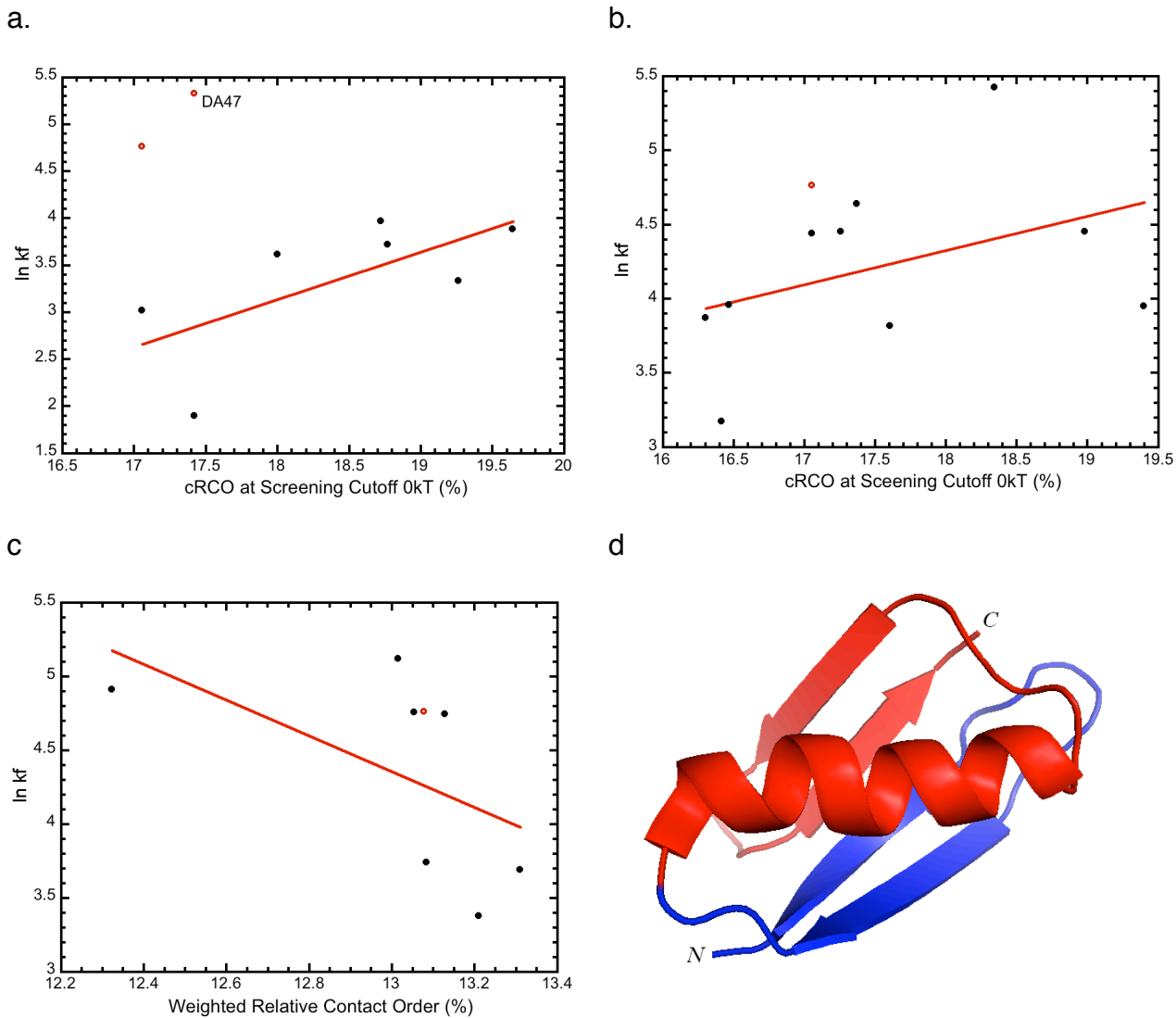

**Figure 5**. Mutant analysis on protein G. (a) C-terminal hairpin of protein G (residues 45-54, R=0.67). The data for DA47 mutation is not taken into consideration, since it involves the removal of a salt bridge, and possible significant change in structure. (b) The helix of protein G (residues 25-35, R=0.41). (c) N-terminal hairpin of protein G (residues 3-18, R=0.55). (d) The color cartoon of protein G shows the early formation of the helix and C-terminal hairpin.

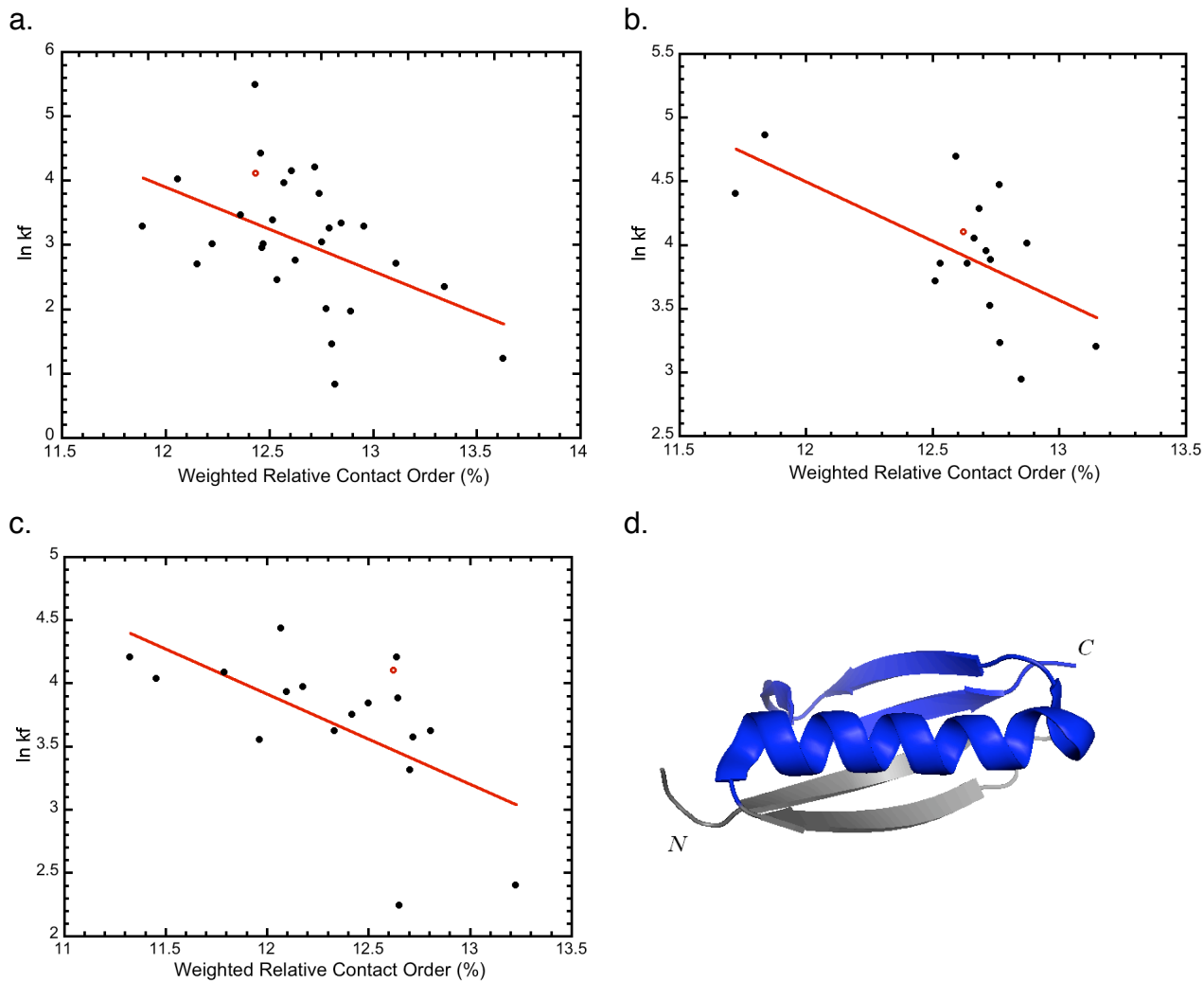

**Figure 6.** Mutant analysis on protein L. (a) N-terminal hairpin of protein L (residues 11-24, R=0.47). (b) The helix of protein L (residues 26-39, R=0.62). (c) C-terminal hairpin of protein L (residues 48-62, R=0.61). (d) The color cartoon of protein L shows that folding rates of mutants at C-terminal hairpin and the helix are highly correlated with wRCO.